# Automated Detection of Acute Lymphoblastic Leukemia Subtypes from Microscopic Blood Smear Images using Deep Neural Networks

Submitted By
**Md. Taufiqul Haque Khan Tusar**
**181472099**
**Roban Khan Anik**
**182482057**

A project report submitted in partial fulfillment of the requirements for the Degree of Bachelor of Science in Computer Science and Engineering

Supervised By
**Md. Touhidul Islam**
Lecturer
Department of CSE
City University

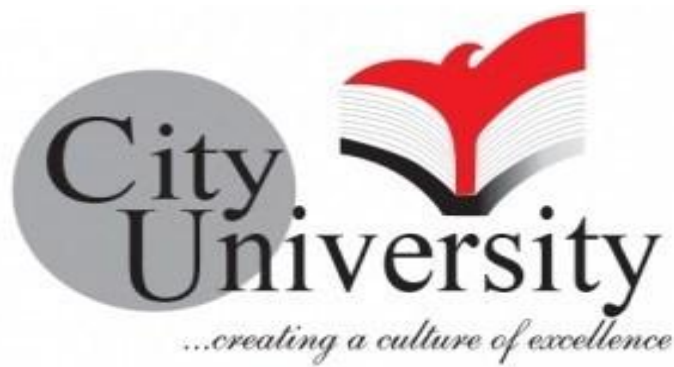

**CITY UNIVERSITY**
**DHAKA, BANGLADESH**
**JULY 2022**

# DECLARATION

We declare that this project report titled "**Automated Detection of Acute Lymphoblastic Leukemia Subtypes from Microscopic Blood Smear Images using Deep Neural Networks**" is the result of our own research except as cited in the references. This project has been done by us under the supervision of **Md. Touhidul Islam.** This project is the partial fulfillment of requirements for the degree awarded as Bachelor of Computer Science and Engineering during the session of 2018-2022 in City University, Dhaka, Bangladesh.

**Submitted By**

*TaufiqKhan*

**Md. Taufiqul Haque Khan Tusar**
ID: 181472099
Batch: 47
Dept. of CSE
City University,
Dhaka, Bangladesh

**Approved By**

*Md. Touhidul Islam*

**Supervisor**

**Md. Touhidul Islam**
Lecturer
Department of Computer Science and Engineering (CSE)
City University, Dhaka, Bangladesh

We declare that this project report titled **"Automated Detection of Acute Lymphoblastic Leukemia Subtypes from Microscopic Blood Smear Images using Deep Neural Networks"** is the result of our own research except as cited in the references. This project has been done by us under the supervision of **Md. Touhidul Islam.** This project is the partial fulfillment of requirements for the degree awarded as Bachelor of Computer Science and Engineering during the session of 2018-2022 in City University, Dhaka, Bangladesh.

**Submitted By**

*Roban Khan*

**Roban Khan Anik**
ID:182482057
Batch: 48th
Dept. of CSE
City University,
Dhaka, Bangladesh

**Approved By**

*Md. Touhidul Islam*

**Supervisor**

**Md. Touhidul Islam**
Lecturer
Department of Computer Science and Engineering (CSE)
City University, Dhaka, Bangladesh

# ACKNOWLEDGEMENT

First, we express our heartiest thanks and gratefulness to almighty Allah for his divine blessing in making us possible to complete this project successfully.

We would like to thank the following people for their help in the production of this project. We are deeply indebted to our supervisor **Md. Touhidul Islam, Lecturer**, Department of CSE. Without his help and support, throughout this project, it would not have been possible. His endless patience, scholarly guidance, continual encouragement, constant and energetic supervision, constructive criticism, valuable advice, and reading many inferior drafts and correcting them at all stages have made it possible to complete this project.

Our special thanks go to the Head of the Department of CSE, **Md. Safaet Hossain**, who had permitted us and encouraged us to go ahead. We are bound to the Honorable Dean of the Faculty of Science and Engineering, **Prof. Dr. Engr. Md. Huamaun Kabir**, for his endless support.

I am very grateful to all my faculty teachers who gave me their valuable guides to complete my graduation. I am also very grateful to all those people who have helped me to complete my project.

Finally, I must acknowledge with due respect the constant support and patients of my parents.

**Md. Taufiqul Haque Khan Tusar**
ID: 181472099
Batch: 47
Dept. of CSE
City University,
Dhaka, Bangladesh

# ACKNOWLEDGEMENT

First, we express our heartiest thanks and gratefulness to almighty Allah for His divine blessing in making us possible to complete this project successfully.
We would like to thank the following people for their help in the production of this project. We are deeply indebted to our supervisor **Md. Touhidul Islam**, Lecturer, Department of CSE. Without his help and support, throughout this project, it would not have been possible. His endless patience, scholarly guidance, continual encouragement, constant and energetic supervision, constructive criticism, valuable advice, and reading many inferior drafts and correcting them at all stages have made it possible to complete this project.

Our special thanks go to the Head of the Department of CSE, **Md. Safaet Hossain**, who had permitted us and encouraged us to go ahead. We are bound to the honorable Dean of the Faculty of Science and Engineering, **Prof. Dr. Engr. Md. Huamaun Kabir**, for his endless support.

I am very grateful to all my faculty teachers who gave me their valuable guides to complete my graduation. I am also very grateful to all those people who have helped me to complete my project.

Finally, I must acknowledge with due respect the constant support and patients of my parents.

**Roban Khan Anik**
ID:182482057
Batch: 48
Dept. of CSE
City University,
Dhaka, Bangladesh

# Abstract

An estimated 300,000 new cases of leukemia are diagnosed each year which is 2.8 percent of all new cancer cases and the prevalence is rising day by day. The most dangerous and deadly type of leukemia is acute lymphoblastic leukemia (ALL) which affects people of all age groups including both children and adults. In this study, we propose an automated system to detect various-shaped ALL blast cells from microscopic blood smears images using Deep Neural Networks (DNN). The system can detect multiple subtypes of ALL cells with an accuracy of 98%. Moreover, we have developed a telediagnosis software to provide real-time support to diagnose ALL subtypes from microscopic blood smears images.

# Table of Contents

# List of Figures

# List of Tables

# Chapter 1

## Introduction

### 1.1 Composition of Blood

Blood consists of several elements. The major components of blood include plasma, red blood cells, white blood cells, and platelets. Plasma is the major constituent of blood and comprises about 55 percent of blood volume. It consists of water with several different substances dissolved within [1]. Almost 45 percent of blood includes Red Blood Cells (RBC), White Blood Cells (WBC), and platelets. The RBC rate ranges from 4,000,000 to 6,000,000 per microliter of blood, representing 40–45% of the total blood volume [2]. WBCs are the cells that defend against germs and give us immunity and resistance; they range from 4500 to 11,000 per microliter of blood [3]. The platelets range from 150,000 to 450,000 per microliter of blood and are responsible for blood clotting [4]. Thus, an increase or decrease in any of the basic blood components will cause problems to a person's health, such as leukemia, thalassemia, and anemia.

### 1.2 Problem Statement

Leukemia is one of the most dangerous types of malignancies affecting the bone marrow or blood in all age groups, both in children and adults. The most dangerous and deadly type of leukemia is acute lymphoblastic leukemia (ALL). Acute Lymphoblastic Leukemia (ALL) is a cancer of the blood cells that are characterized by a large number of immature lymphocytes, known as blast cells (myeloblasts) [5].  lymphocytes are classified into three types. These are (I) normal lymphocytes, (II) atypical lymphocytes, and (III) reactive lymphocytes. Normal lymphocytes are characterized by homogeneity and round, small, and rough nuclei. Atypical cells, by a large size and nucleus and the fact that they have lumpy chromatins. Reactive cells, by their heterogeneity and the fact that they are surrounded by red cells. Microscopic examination is the method of diagnosing lymphocyte types. It involves taking blood or bone marrow samples, which are diagnosed by a pathologist [6]. ALL is classified into three morphological types: (I) L1 acute lymphoblastic leukemia, (II) L2 acute lymphoblastic leukemia, and (III) L3 acute lymphoblastic leukemia [7]. There, L1 cells are the smallest, with a uniform population and coarse chromatins. L2 cells have nuclear heterogeneity and are larger than L1 cells. L3 cells have vacuoles protruding into the cells and are larger than L1 cells [8]. About 90 % of patients can be cured, provided that the disease is diagnosed at an early stage by adopting mass screening procedures with an intelligent automated diagnosis system.

### 1.3 Prevalence of Leukemia

According to the statistics from the World Health Organization's (WHO) International Agency for Research on Cancer (IARC), the number of leukemia cases in both sexes of all age groups in 2018 was 4,37,033 and a total of 3,03,006 deaths [9]. Globally, the incidence rate (per 1,00,000) was 5.2 and the mortality rate (per 1,00,000) was 3.5. A study was conducted from January 2014 to December 2016 in the department of Pediatric Hematology and oncology of Dhaka Shishu Hospital, Bangladesh. During this period all diagnosed children with ALL (12 months to 16 years) were included in the study who were selected to give chemotherapy. Out of the total of 597 children, 327 (54.7%) were males and 270 (45.3%) were female. Parents of a total of 199 (3.33%) children refused to start therapy and the rest 398 children began chemotherapy. Among the total children who started chemotherapy, 120 (20.10%) abandoned treatment in different phases of therapy. 66 (55%) abandoned treatment during induction of remission. 23 (19.16%) during consolidation, 12(10%) re-induction, and 19 (1585%) children during maintenance phases of chemotherapy [10]. There the 199 patient's parents refused treatment because of the high cost and the treatment duration was very lengthy (about 10 years). The study shows the severity of ALL.

## 1.4 Difficulties in Existing Manual Diagnosis

The evaluation of bone marrow morphology and blood smear in an electronic microscope by experienced hematopathologists is essential in the diagnosis of Acute Lymphoblastic Leukemia (ALL). However, it suffers from a lack of standardization and inter-observer variability. Because diagnosing Acute Lymphoblastic Leukemia (ALL) is a tedious task that is prone to human and machine errors. In several instances, it is difficult to make an accurate final decision even after careful examination by an experienced pathologist. Manual detection of ALL cells is conducted by pathologists and is typically subject to human error and produces inaccurate results. This process is tedious, time-consuming, and subject to inter and intra-class variations among pathologists. Only 76.6% of the cases showed agreement between pathologists during leukemia diagnosis [11].

## 1.5 Challenges in Automated Diagnosis using DNN

The challenges in automated detection are associated with the complex nature of WBCs, including irregular boundaries and the textural similarities between WBCs and other blood components, which cause difficulties in separating blast cells (cancer cells) and healthy cells. [4] Besides The Learnability of a Deep Learning Model strongly depends on the effective and efficient data preparation pipeline. The accuracy and time complexity of ALL detection is strongly associated with the quality of the extracted features used in training the pixel-wise classification models.

## 1.6 Proposed Approach

In this thesis, we propose to develop multiple robust Deep Neural Network Models for Medical Image Data with Acute Lymphoblastic Leukemia (ALL) cells, with enhanced learnability, predictability, and accuracy. Besides, we develop a Web Application to provide real-time support to diagnose ALL cells. We have also analyzed the performance of our proposed method with multiple related works to assure the robustness of the proposed model.

## 1.7 Contributions of the Proposed Approach

- Subtype classification of Acute Lymphoblastic Leukemia with almost 98% accuracy.
- Robust Deep Neural Network model for multiclass imbalanced data sets.
- Real-time telediagnosis support through the web application which has been built based on the proposed approach.

## 1.8 Organization of the Study

We organize the study as follows. In Chapter 2 Literature Review, we briefly describe the related work's methodology, outcomes, and limitations. In Chapter 3 Dataset Description, we elaborately discuss the data set and illustrate the image samples. After that in Chapter 4, the proposed method is detailed. Chapter 5 Result and Discussion elaborate on the performance of the models. Finally, In Chapter 6 we discussed the developed telediagnosis software. In the end we have concluded the research and mentioned the future works.

# Chapter 2

# Literature Review

Researchers applied various methods which incorporate Deep Learning (DL), Deep Neural Networks (DNN), and Medical Image Processing to detect leukemia cells from blood smear images and bone marrow images.

Rohan et. al [5] developed an automation system to detect ALL blast cells in microscopic blood smears using the You Only Look Once (YOLO v4) algorithm. They trained and evaluated their method using the ALL-IDB1 and C_NMC_2019 data sets. The ALL-IDB1 dataset contained 108 images, which were collected in September 2005. The C_NMC_2019 dataset contained 10,661 images collected from 73 subjects. The images in this dataset were single-celled and these were already pre-processed. In the preprocessing step they applied Auto Orientation, Resizing, and Augmentation techniques. The preprocessing steps generate 872 pre-process images after augmentation on the ALL-IDB1 dataset. Finally achieved The mAP (Mean Average Precision) of 96.06 % for the ALL-IDB1 dataset and 98.7 % for the C_NMC_2019 dataset. Where the loss was seen to reduce exponentially and reached an average of 0.57664 at the end of 6000 iterations on the ALL-IDB1 dataset and also the loss was seen to reduce exponentially and reached an average of 0.502 at the end of 6000 iterations on the C_NMC_2019 dataset. But no complete automated screening or diagnosing telehealth facility was established in this study.

The authors in [8] proposed 3 distinct systems which were built based on ALL-IDB1 and ALL-IDB2. The first consists of the artificial neural network (ANN), feed-forward neural network (FFNN), and support vector machine (SVM), all of which are based on hybrid features extracted using Local Binary Pattern (LBP), Gray Level Co-occurrence Matrix (GLCM) and Fuzzy Color Histogram (FCH) methods. The second proposed system consists of the convolutional neural network (CNN) models: AlexNet, GoogleNet, and ResNet-18, based on the transfer learning method, in which deep feature maps were extracted and classified. The third proposed system consists of hybrid CNN–SVM technologies, consisting of two blocks: CNN models for extracting feature maps and the SVM algorithm for classifying feature maps. In the first system, ANN and FFNN reached an accuracy of 100%, while SVM reached an accuracy of 98.11%. In the second proposed system, all the transfer learning models have achieved 100% accuracy and in the last proposed system AlexNet + SVM achieved 100% accuracy, Goog-LeNet + SVM achieved 98.1% accuracy, and ResNet-18 + SVM achieved 100% accuracy. Although the feature extraction methods are well enough, the data preparation pipeline is not generalized and effective because of the single Denoising technique (average and Laplacian filters) in the pre-processing step. Where the robustness of the model comes into question.

A multi-step DL approach was performed in [12] to automatically segment Leukemia cells from bone marrow images. The authors analyzed 1251 patients who have been newly diagnosed and treated with AML and 236 Healthy Cohort. Then build the model using 5204 augmented AML bone marrow smear images and 5428 augmented control bone marrow smear images. The multi-step includes the following workflow. First, initial segmentation was done with the VGG Image Annotator tool. Then, the FRCNN was trained with the segmented images and created new segmentation proposals for unsegmented images which were manually corrected by hematologists. Step 2: Feature extraction was performed manually by hematologists. Step 3: For the distinction between AML and healthy control samples based on segmented images, the authors trained a multitude of DL models for binary predictions. In step 4: For NPM1 status prediction, transfer learning with a ResNet50 pre-trained on ImageNet was utilized on BMS images. Finally, the FRCNN achieved a cell segmentation accuracy of 0.97 from BMS. The binary classification model showed an AUC of 0.97 for both the ROC and the precision-recall curve and a micro-average accuracy of 0.91 distinguishing between AML and healthy bone marrow donor samples. For NPM1 prediction their DL model achieved a high accuracy of 0.86 in predicting mutation status. Although the study has a good outcome, the manual feature extraction technique may lead to inefficiency and error with the large dataset.

The AML_Cytomorphology_LMU dataset consisted of 18,365 expert-labeled single-cell images obtained from peripheral blood smears of 100 AML patients and 100 controls at Munich University Hospital between 2014 and 2017 used by the study [13]. The author proposed a new hybrid feature extraction method using image processing and deep learning methods. The proposed method consists of two steps: 1) a region of interest (ROI) is extracted using the CMYK-moment localization method and 2) deep learning-based features are extracted using a CNN-based feature fusion method. The method obtained overall classification accuracies of 97.57%. There the image data preparation pipeline is not generalized which can lead to a mass accuracy fall in terms of new image data.

Authors developed a method in [26] to detect ALL using transfer learning and later validate the model using XAI. As the dataset was imbalanced with the more ALL class, the class weight method was used to balance the weights of the two classes in preprocessing For the used dataset, the class weight of normal cells was 1.57288, and the class weight of acute lymphocytic leukemia (ALL) cells was 0.73301. Then, different pre-trained models, namely InceptionV3, ResNet101V2, VGG19, and InceptionResNetV2, were used to train the model. All the images were reduced to 299px * 299px because the Inception V3 input shape must be (299, 299, 3) and a standard batch size of 32 is used while training. A segmentation method was employed to divide the example image into separate sections to see if the model could accurately read it. ResNet101v2 model Accuracy is 0.9861, Loss is 0.0333 and validation accuracy is 0.9589, Validation loss is 0.1559 F1 score 0.9861, and Validation F1 score 0.9588.

The author of this paper [27] established a database, which consists of 1,732 images obtained from the bone marrow smears of 89 children with leukemia from 2009 to 2019 at the Shanghai Children's Medical Center (SCMC). For multi-class WBC differentiation, they developed three deep learning models (ResNext101_32∗8d swsl, ResNext50_32∗4dswsl, and ResNet50). Finally, the Accuracy, Ap, F1 score, and AUC are 0.8149, 0.7982, 0.8073, and 0.8293 respectively. In this study, authors retrospectively collected 1,732 bone marrow images containing 27,184 cells (including 24,165 cells and 2,983 cell debris) from 89 children with leukemia from the Shanghai Children's Medical Center. We randomly separated 70% of the cells in the training set, and the remaining cells were used to form the validation set and test set.

In this study [28] 4,451 real images of bone marrow cells on the Hyde star HDS-BFS high-speed micro scanning image system were collected, with a resolution of 4,000×3,000. The object detection model is based on the Faster RegionConvolutional Neural Network (R-CNN). Faster R-CNN is composed of the region proposal network (RPN) and Fast R-CNN, in which RPN is used to select candidate target boxes and Fast R-CNN is used for accurate target classification and regression. Compared to the baseline, although their method resulted in a slight drop in recall (~4%), it improved the precision by 26.4%, the F1-score by 12.1%, and the AP@50 by 3%. The model achieves a recall of 0.710, precision of 0.496, AP@50 of 0.533, and F1-score of 0.575. They developed a new morphological diagnosis system for bone marrow cells. The model is based on the Faster R-CNN.

The study [29] focuses on the C-NMC-2019 dataset and the application of DL and the ensemble strategy. In their proposed model, the Hem class was over-sampled to 5822 images, and a total of 11644 images were trained during the training process. In this circumstance, five pre-trained networks, VGG-16, Xception, MobileNetV2, InceptionResNet-V2, and DenseNet-121, are adopted for transfer learning applications. The image is transferred through a stack of convolutional layers, where the filters are used with tiny receptive filters ($3 \times 3$). The Kappa-based ensemble model has produced the best ALL recognition results, with 89.72% WFS and 94.8% AUC.

The authors in [31], applied feature extraction using PCA 2 to reduce dimensionality in a heart disease dataset. There the proposed method integrates the K-means Clustering and Genetic Algorithm and finally achieved 94.06% of clustering accuracy. In [32], the authors proposed a novel feature selection algorithm in a chronic kidney disease dataset that aggregated wrapper, filter, and ensemble methods and achieved 100% accuracy in prediction.

# Chapter 3

## Dataset Description

We have used the Acute Lymphoblastic Leukemia (ALL) Image dataset. The images of this dataset [24] [25] were prepared in the Bone Marrow Laboratory of Taleqani Hospital (Tehran, Iran). This dataset consisted of 3256 peripheral blood smear (PBS) images from 89 patients suspected of ALL, whose blood samples were prepared and stained by skillful laboratory staff, including 25 healthy individuals with a benign diagnosis (hematogone) and 64 patients with a definitive diagnosis of ALL subtypes, Early Pre-B, Pre-B, and Pro-B ALL. This dataset is divided into two classes benign and malignant. The former comprises hematogenous and the latter is the ALL group with three subtypes of malignant lymphoblasts: Early Pre-B, Pre-B, and Pro-B ALL. All the images were taken by using a Zeiss camera in a microscope with a 100x magnification and saved as JPG files. A specialist using the flow cytometry tool made the definitive determination of the types and subtypes of these cells. After color thresholding-based segmentation in the HSV color space, dataset collectors also provide segmented images. Figure 3.1.1 - 3.1.4 depicts the images of Benign (hematogone) Cells, Early Pre-B ALL Cells, Pre-B ALL Cells, and  Pro-B ALL Cells respectively.

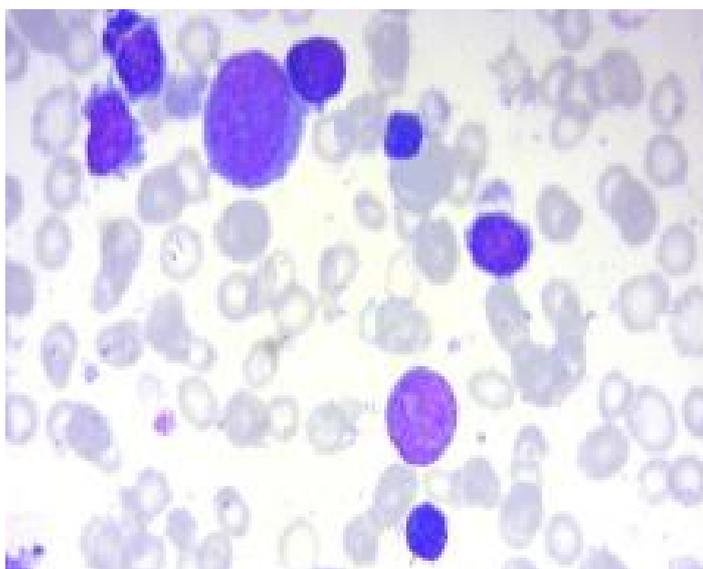

**Fig. 3.1.1** Benign (hematogone) cells

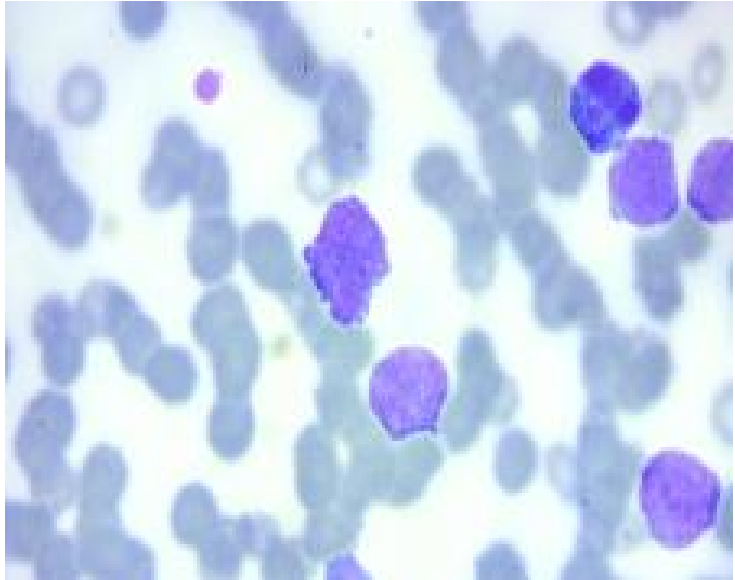

**Fig. 3.1.2** Early Pre-B ALL cells

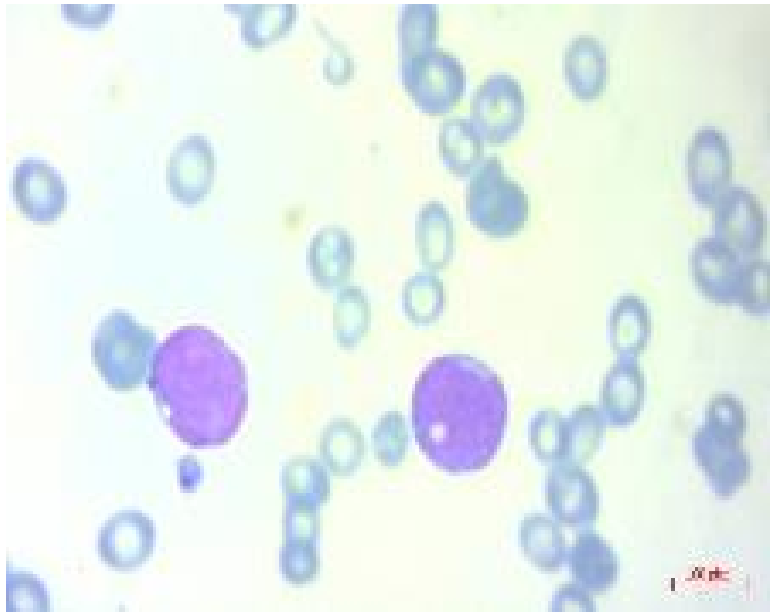

**Fig. 3.1.3** Pre-B ALL cells

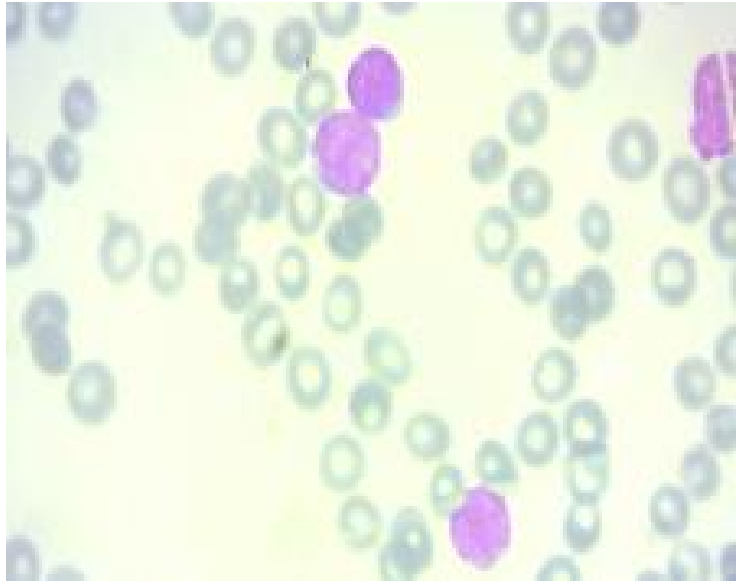

**Fig. 3.1.4** Pro-B ALL cells

We have used the Acute Lymphoblastic Leukemia (ALL) Dataset to train and validate our DNN model. It is a multiclass dataset that contains a benign cell and 3 subtypes of blast cells named Early Pre-B, Pre-B, and Pro-B. A DNN model trained with ALL datasets may aid in detecting the subtypes of Acute Lymphoblastic Leukemia by classifying all types of subtypes.

# Chapter 4

# Proposed Method

## 4.1 Overview

We have selected the Acute Lymphoblastic Leukemia (ALL) public datasets to build our Deep Neural Network models. Figure 4.1.1 illustrates the workflow of the proposed method.

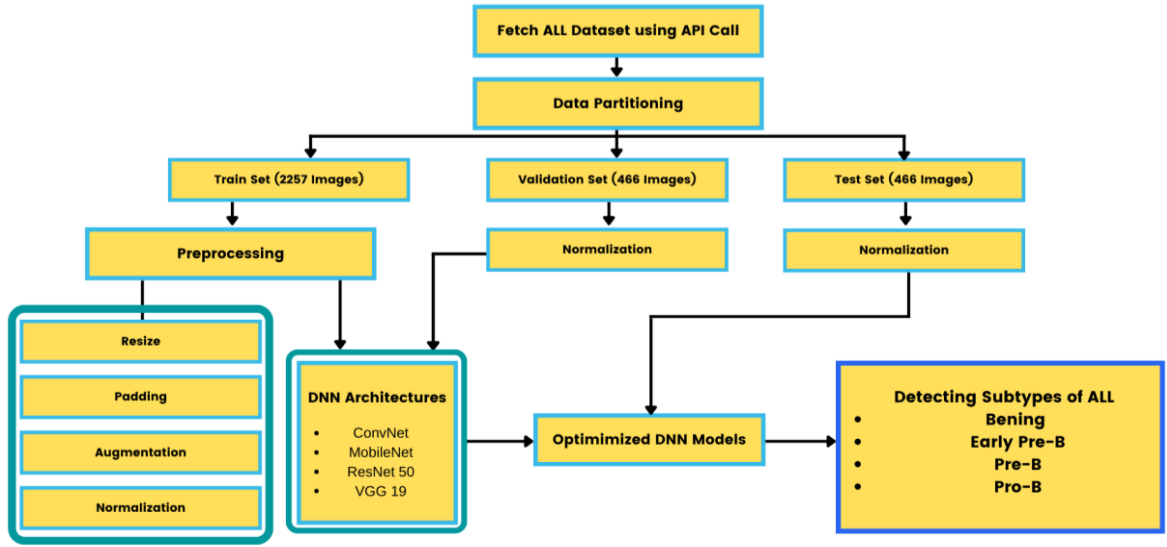

**Fig. 4.1.1** Workflow of the proposed method

In the beginning, we called the Kaggle API to fetch the dataset. Then we split the dataset into Training, Validation, and Testing Set to avoid data leakage and overfitting. Then we preprocessed the Training Set using our Data Preprocessing Pipeline. Then we have only normalized the Validation and Testing Set's images. It will keep the Global and Local Minima-Maxima different, which is another essential procedure to avoid data leakage and overfitting. Then the Training and Validation images have passed to the DNN architectures. Now, Multiple DNN Algorithms have been trained and validated from the Training and Validation Set. The best-fitted models depending on validation accuracy at this stage have been selected as optimized models. Now the normalized Test images have been tested with the Multiple Optimized DNN Models. Finally, the images in the Test set will be classified into Benign, Early Pre-B, Pre-B, and Pro-B to detect various subtypes of acute lymphoblastic leukemia cells.

## 4.2 Data Preparation Pipeline

The performance of a DNN depends largely on suitable data preparation. Our data preparation pipeline contains the following techniques to pre-process the microscopic images for ALL detection.

- Resizing (224px width, 224px height, 3 channels)
- Padding
- Augmentation
- Normalization

There the Augmentation technique contains the following techniques.

- Sheared Range ( 0.2)
- Zooming (0.2)
- Flipping (Horizontally)

## 4.3 Deep Neural Network Architecture

We have implemented the following DNN algorithms for our thesis work.
- Convolutional Neural Network (ConvNet)
- MobileNetV2
- Residual Neural Network 50 (ResNet50)
- Visual Geometry Group 2019 (VGG19)

### 4.3.1 Applied Convolutional Neural Network (ConvNet)

We have built an 11-layer Convolutional Neural Network with 4 Convolutional Layers, 3 Max Pooling layers, 2 Dropout layers, a single Flatten, and a Dense layer with SoftMax activation function for multi-class classification. The ConvNet model contains 5,638,440 trainable parameters.

### 4.3.2 Applied MobileNetV2 Architecture

We have built a MobileNetV2 architecture with 81 layers. There are 13 Conv2D layers, 24 Convolutional layers for Batch Normalization, 26 layers for Relu activation function, 12 Depth wise Convolutional layers, 4 Zero padding layers, a single Flatten layer, and finally a Dense layers with 4 unit and SoftMax activation function. There are total parameters of 3,429,572 where 200,708 parameters are trainable and 3,228,864 non-trainable parameters.

### 4.3.3 Residual Neural Networks 50 (ResNet50)

We have built a ResNet50 with 149 layers, where 42 Activation layers, 46 Convolutional Layers, 45 layers for Batch wise Normalization, 16 Add layers, 2 Zero Padding layers, a Flatten layer, and a Dense layer. There is a total of 23,989,124 parameters, 401,412 trainable parameters, and 23,587,712 Non-trainable parameters.

### 4.3.4 Visual Geometry Group 2019 (VGG19)

The VGG19 architecture contains a total of 20,124,740 parameters, where 100,356 parameters are trainable and 20,024,384 non-trainable parameters. There are 16 Convolutional layers, 5 Max Pooling layers, a single Flatten layer, and a single Dense layer.

The complete code of the study is available in the following GitHub repository.
https://github.com/Muhammad-Taufiq-Khan/Automated-Detection-of-Acute-Lymphoblastic-Leukemia-Subtypes-from-Microscopic-Blood-Smear-Images

# Chapter 5

## Results & Discussion

### 5.1 MobileNetV2 Model

The Optimized MobileNetV2 Model provides Training, Validation, and Testing accuracy of 0.9312, 0.9799, and 0.9742 respectively. There the Training, Validation, and Testing Loss are 0.7753, 0.1792, and 0.2351 respectively. Figure 5.1.1 shows the result of the MobileNetV2 Model. Figure 5.1.2 shows the Training accuracy vs validation accuracy of the MobileNetV2 Model and Fig. 5.1.3 shows the Training loss vs Validation loss of the MobileNetV2 Model.

|  | Accuracy | Loss |
|---|---|---|
| Training | 0.9312 | 0.7753 |
| Validation | 0.9799 | 0.1792 |
| Testing | 0.9742 | 0.2351 |

**Fig. 5.1.1** Result of the MobileNetV2 model

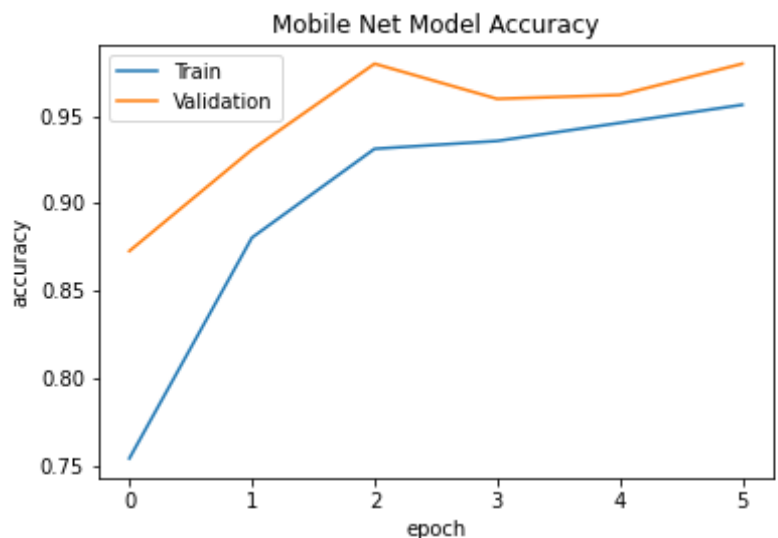

**Fig. 5.1.2** Training accuracy vs validation accuracy of the MobileNetV2 model

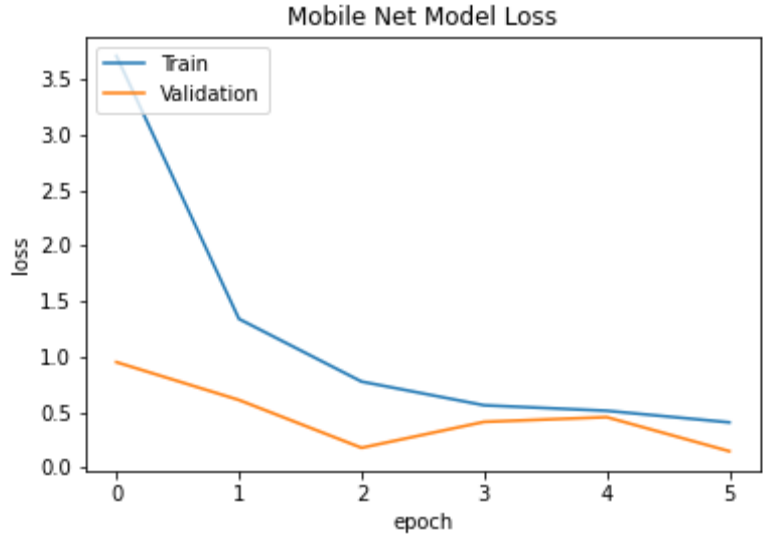

Fig. **5.1.3** Training loss vs validation loss of the MobileNetV2 model

## 5.2 Convolutional Neural Network (ConvNet) Model

The accuracy and loss of the Optimized ConvNet Model in the Training, Validation, and Testing has been shown in Fig. 5.2.1. Figure 5.1.2 shows the Training accuracy vs Validation accuracy of the model and Fig. 5.1.3 shows the Training loss vs validation loss of the model.

|            | Accuracy | Loss   |
|------------|----------|--------|
| Training   | 0.8939   | 0.2799 |
| Validation | 0.9643   | 0.1282 |
| Testing    | 0.9128   | 0.2309 |

**Fig. 5.2.1** Result of the ConvNet model

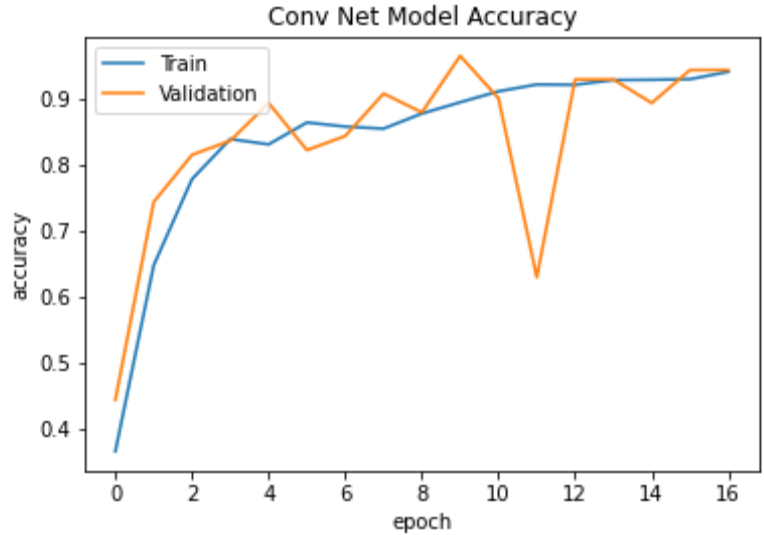

**Fig. 5.2.2** Training accuracy vs validation accuracy of the ConNet model

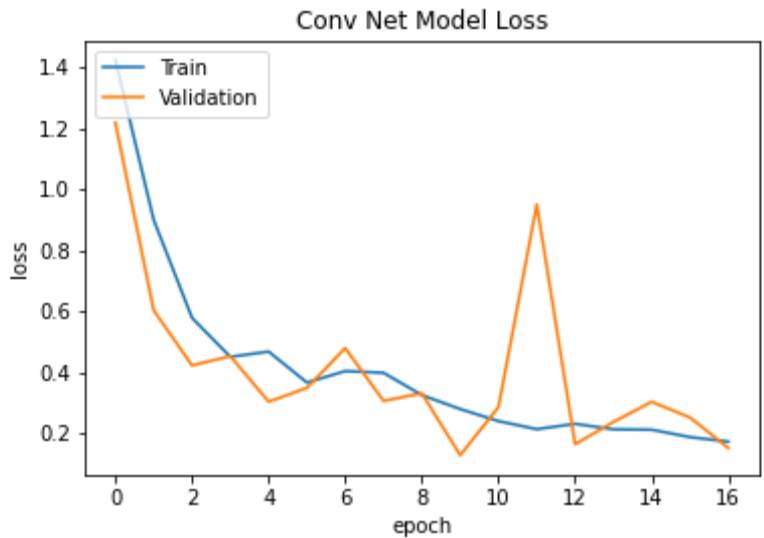

**Fig. 5.2.3** Training loss vs validation loss of the ConvNet model

## 5.3 Residual Neural Network 50 (ResNet 50) Model

The accuracy and loss of the Optimized ResNet Model on the Training, Validation, and Testing has been shown in Fig. 5.3.1. Figure 5.3.2 shows the Training accuracy vs validation accuracy of the model, and Fig. 5.3.3 shows the Training loss vs validation loss of the model.

| | Accuracy | Loss |
|---|---|---|
| Training | 0.7151 | 2.4317 |
| Validation | 0.8237 | 1.0540 |
| Testing | 0.8526 | 0.8412 |

**Fig. 5.3.1** Result of the ResNet50 model

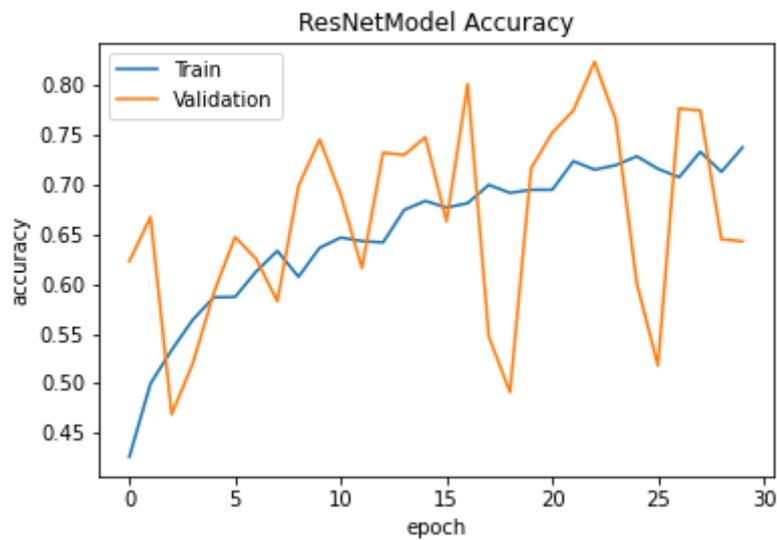

**Fig. 5.3.2** Training accuracy vs validation accuracy of the ResNet50 model

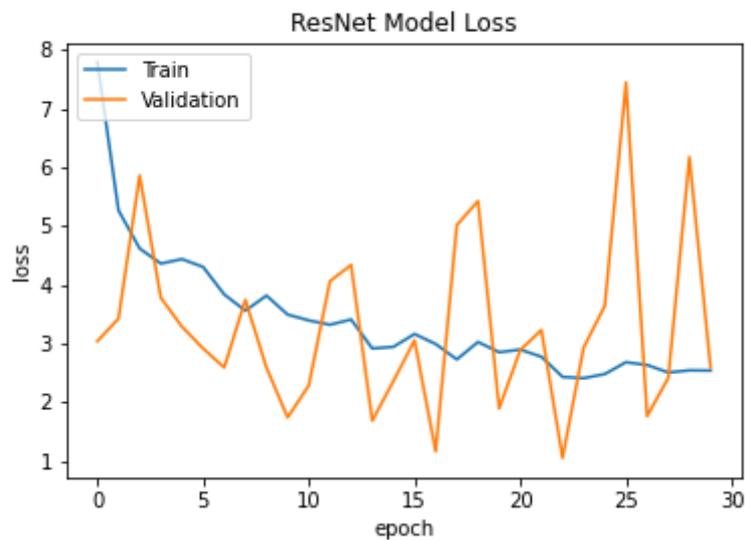

**Fig. 5.3.3** Training loss vs validation loss of the ResNet50 model

## 5.4 Visual Geometry Group 2019 (VGG19) Model

The accuracy and loss of the Optimized VGG19 Model on the Training, Validation, and Testing has been shown in Fig. 5.4.1. Figure 5.4.2 shows the Training accuracy vs validation accuracy of the model, and Fig. 5.4.3 shows the Training loss vs validation loss of the model.

|  | **Accuracy** | **Loss** |
|---|---|---|
| **Training** | 0.9488 | 0.1557 |
| **Validation** | 0.9643 | 0.1184 |
| **Testing** | 0.9613 | 0.099 |

**Fig. 5.4.1** Result of the VGG19 model

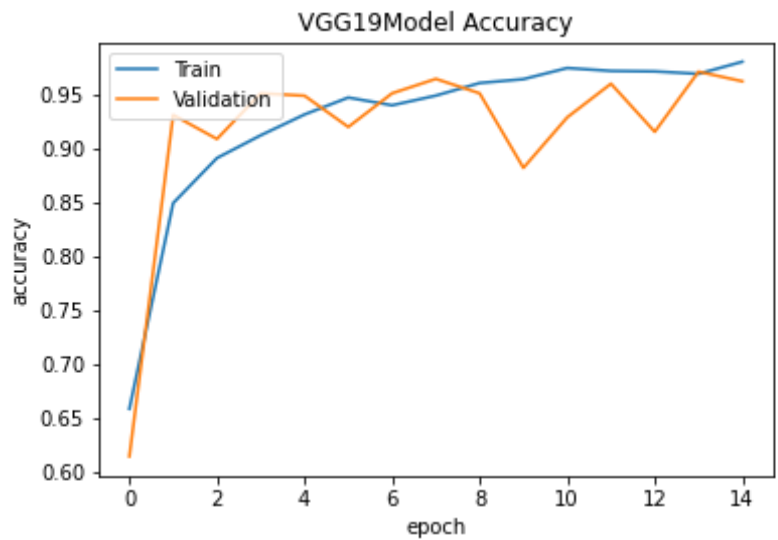

**Fig. 5.4.2** Training accuracy vs validation accuracy of the VGG19 model

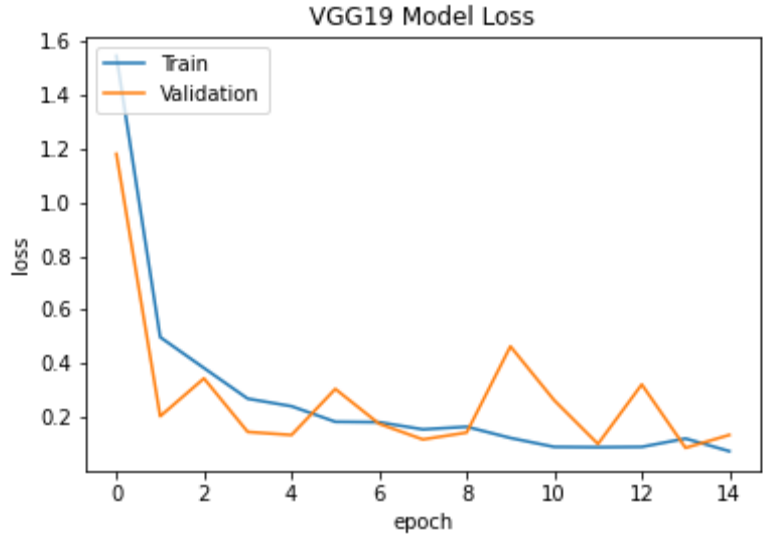

**Fig. 5.4.3** Training loss vs validation loss of the VGG19 model

## 5.6 Comparative Analysis with Related Works

TABLE I
TESTING ACCURACY AND LOSS OF THE RELATED WORKS

| Title & Year | Algorithm | Accuracy | Loss |
|---|---|---|---|
| Automated blast cell detection for Acute Lymphoblastic Leukemia diagnosis (2021) [5] | YOLOv4 for ALL_IDB1<br><br>YOLOv4 for C_NMC-2019 | 92.0%<br><br>96.0% | 8.0%<br><br>4.0% |
| Feature Extraction of White Blood Cells Using CMYK -Moment localization andDeep Learning in Acute Myeloid Leukemia Blood Smear Microscopic Image (2022) [13] | FCL<br><br>RF<br><br>SVM<br><br>XGBoost | 94.92%<br><br>95.47%<br><br>96.41%<br><br>95.18% | 5.08%<br><br>4.53%<br><br>3.59%<br><br>4.82% |
| Executing Spark BigDl for Leukemia Detection from Microscopic Image using Transfer Learning (2021) [30] | GoogleNet<br><br>CNN | 96.06%<br><br>94.69% | 3.995%<br><br>5.32% |

| Hybrid Inception v3 XGBoost Model for Acute lymphoblastic Leukemia Classification (2021) [7] | AlexNet | 89.4% | 10.6% |
| | DenseNet121 | 86.9% | 13.1% |
| | ResNet18 | 91.7% | 8.3% |
| | VGG16 | 92.4% | 7.65% |
| | SqueezeNet | 93.2% | 6.8% |
| | MobileNetV2 | 95.8% | 4.2% |
| Development and Evaluation of a Leukemia Diagnosis System Using Deep Learning in Real Clinical Scenarios (2021) [27] | ResNet101 | 81.49% | 18.51% |
| | ResNext5 | 79.82% | 20.18% |
| | ResNet50 | 80.73% | 19.27% |
| | Ensemble | 82.93% | 17.07% |

## 5.7 Discussion

Analyzing the above results from Fig. 5.1.1 - 5.4.3 we can conclude that except for ResNet50 the other models provide desirable outcomes. There the MobileNetV2 model performs higher in terms of Training Accuracy and VGG19 provides the lowest loss. Table II shows the Testing accuracy and loss of the optimized models below. Also, compared to Table I the MobileNetV2 model provides higher accuracy.

TABLE II
TESTING ACCURACY AND LOSS OF THE DNN MODELS OF THE PROPOSED METHOD

| | Accuracy | Loss |
|---|---|---|
| MobileNetV2 | **0.9742** | 0.2351 |
| VGG19 | 0.9613 | **0.099** |
| ConvNet | 0.9128 | 0.2309 |
| ResNet50 | 0.8526 | 0.8412 |

# Chapter 6

## Telediagnosis Software Based on the Proposed Method

### 6.1 Telediagnosis Service of the WebApp

We have developed a Realtime Tele Diagnosis Web App to aid in the detection of ALL subtypes. Fig 6.1.1 illustrates the architecture of the Telediagnosis software. The App collects ALL Images from the user-end (Frontend) and passes the image data to the server-end (Backend). The server processes the images and calls the best DNN model. Finally, the model predicts the subtype of the ALL and passes the result to the user-end. Fig 6.1.2 illustrates the home page of the Web Application and fig 6.1.3 illustrates how the diagnosis result is shown to the user-end.

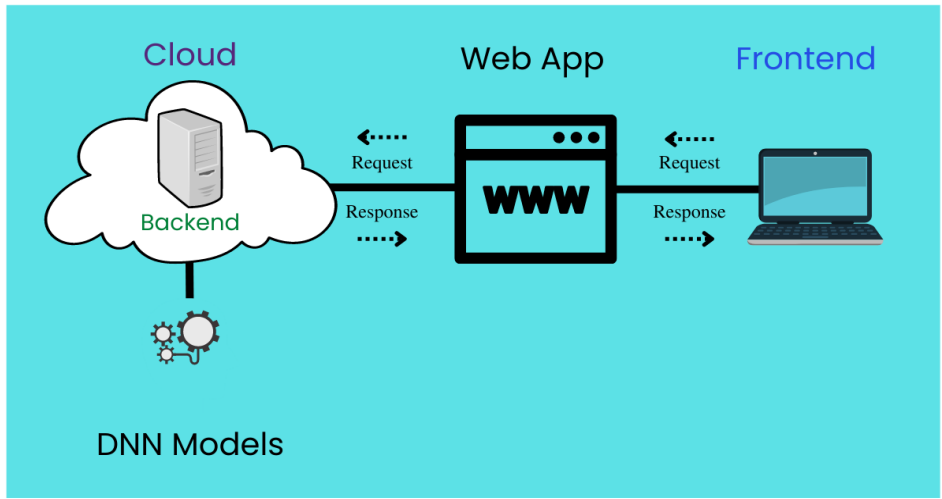

**Fig. 6.1.1** Architecture of the telediagnosis software

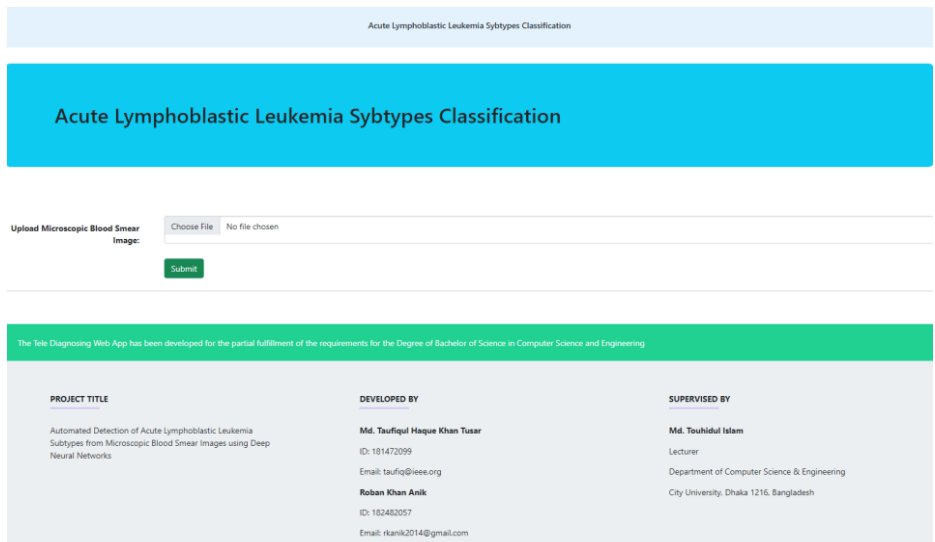

**Fig. 6.1.2** Web Application home page

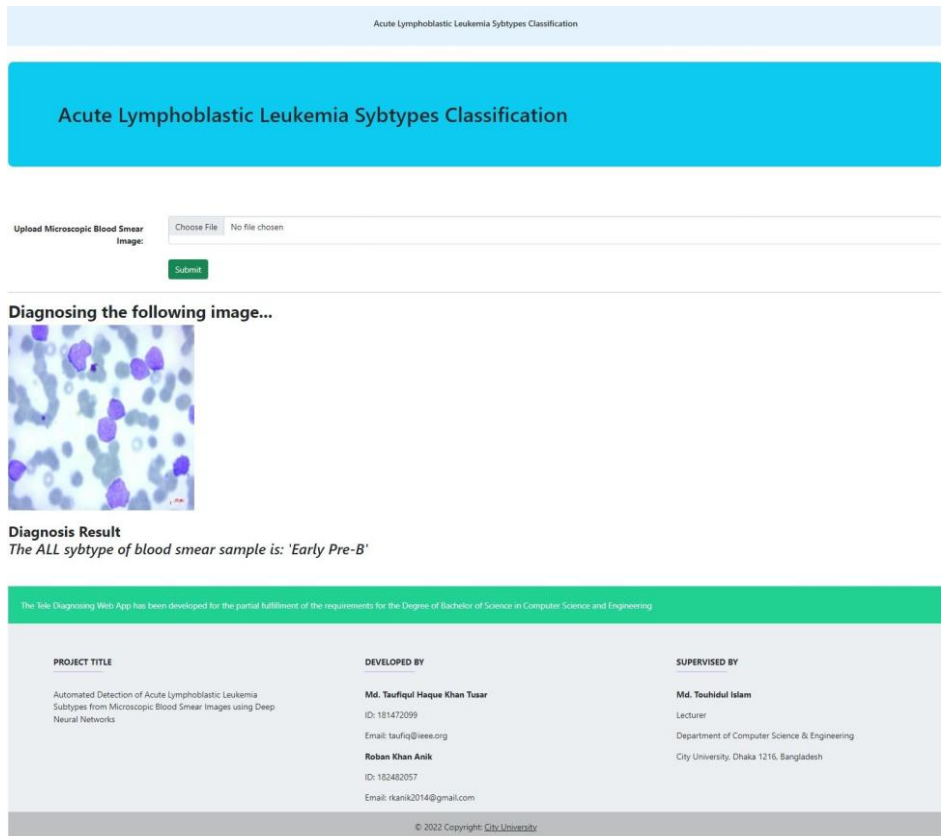

**Fig. 6.1.3** Diagnosing blood smear image

The complete code of the web application is available in the following github repository.
https://github.com/Muhammad-Taufiq-Khan/ALL_Subtype_Detector_WebApp

## 6.2 Web App Development Environments

- Programming Language: Python
- Backend: Flask
- Frontend: Bootstrap5, HTML, CSS
- Deep Learning Library: Tensorflow-Keras
- Digital Image Processing Library: OpenCV
- Cloud Platform: Heroku.

# Conclusion

A lot of people are losing lives because of the severity of Acute Lymphoblastic Leukemia and the number is increasing day by day. The patients who survive lose their vitality. The current Artificial Intelligence-based automated systems are also facing challenges due to the various shape, patterns, and textures of ALL blast cells. We have developed optimized Multi-DNN models and a telediagnosis web application, which will fight against the severity of ALL by providing efficient diagnosis assistance.

# Future Work

According to our research many researchers have applied DNN to detect Acute Lymphoblastic Leukemia but a thorough method and generalized model are still missing. In the future, we would like to develop a robust method, which will be able to perform segmentation, ROI detection, estimated blast cell calculation, and generalized classification.

## List of Abbreviation and Acronyms

| | |
|---|---|
| AML | Acute Myeloid Leukemia |
| ALL | Acute Lymphoblastic Leukemia |
| DNN | Deep Neural Networks |
| NPM1 | Nucleophosmin Gene |
| FRCNN | Faster R-CNN Object Detection Network |
| WHO | World Health Organization |
| VGG19 | Visual Geometry Group 2019 |
| ResNet50 | Residual Neural Network 50 |
| ConvNet | Convolutional Neural Network |
| YOLOV4 | You Only Look Once |
| FCL | Fully Connected Layers |
| RF | Random Forest |
| SVM | Support Vector Machine |
| XGBoost | Extreme Gradient Boosting |
| FFNN | Feed Forward Neural Network |
| FCH | Fuzzy Color Histogram |

| LBP | Local Binary Pattern |
|-----|----------------------|
| PBS | Peripheral Blood Smear |
| ANN | Artificial Neural Network |